\documentclass{article}
\usepackage[utf8]{inputenc}
\usepackage{amsmath, amsthm, amssymb,color,fullpage,url,booktabs}  
\usepackage{graphicx}
\ifdefined\skiphyperref
\else
\ifdefined\blackandwhite
\usepackage[pdftex]{hyperref} 
\else
\usepackage[colorlinks,pdftex]{hyperref} 
\hypersetup{citecolor = blue, linkcolor = red!70!black}
\fi
\fi

\usepackage{cleveref}
\usepackage{tikz}
\usepackage{authblk}

\newcommand{\nc}{\newcommand}
\nc{\rnc}{\renewcommand}

\newcommand{\bra}[1]{\left\langle #1\right|}
\newcommand{\ket}[1]{\left|#1\right\rangle}

\def\be#1\ee{\begin{equation}#1\end{equation}}
\def\ba#1\ea{\begin{align}#1\end{align}}
\def\bas#1\eas{\begin{align}#1\end{align}}
\def\bpm#1\epm{\begin{pmatrix}#1\end{pmatrix}}
\nc{\non}{\nonumber}
\nc{\nn}{\nonumber}
\nc{\eq}[1]{(\ref{eq:#1})}
\nc{\eqs}[2]{(\ref{eq:#1}) and (\ref{eq:#2})}
\nc{\ra}{\rightarrow}
\nc{\ot}{\otimes}
\nc{\grad}{{\vec{\nabla}}}

\def\bea#1\eea{\begin{eqnarray}#1\end{eqnarray}}
\def\beas#1\eeas{\begin{eqnarray*}#1\end{eqnarray*}}

\makeatletter
\newtheorem*{rep@theorem}{\rep@title}
\newcommand{\newreptheorem}[2]{%
\newenvironment{rep#1}[1]{%
 \def\rep@title{#2 \ref{##1} (restatement)}%
 \begin{rep@theorem}}%
 {\end{rep@theorem}}}
\makeatother

\newreptheorem{theorem}{Theorem}
\newreptheorem{lemma}{Lemma}

\nc\eps{\epsilon}

\nc\cA{\mathcal{A}}
\nc\cB{\mathcal{B}}
\nc\cC{\mathcal{C}}
\nc\cD{\mathcal{D}}
\nc\cE{\mathcal{E}}
\nc\cF{\mathcal{F}}
\nc\cG{\mathcal{G}}
\nc\cH{\mathcal{H}}
\nc\cI{\mathcal{I}}
\nc\cJ{\mathcal{J}}
\nc\cK{\mathcal{K}}
\nc\cL{\mathcal{L}}
\nc\cM{\mathcal{M}}
\nc\cN{\mathcal{N}}
\nc\cO{\mathcal{O}}
\nc\cP{\mathcal{P}}
\nc\cQ{\mathcal{Q}}
\nc\cR{\mathcal{R}}
\nc\cS{\mathcal{S}}
\nc\cT{\mathcal{T}}
\nc\cU{\mathcal{U}}
\nc\cV{\mathcal{V}}
\nc\cW{\mathcal{W}}
\nc\cX{\mathcal{X}}
\nc\cY{\mathcal{Y}}
\nc\cZ{\mathcal{Z}}

\nc\bbC{\mathbb{C}}

\nc\bbF{\mathbb{F}}
\nc\bbM{\mathbb{M}}
\nc\bbN{\mathbb{N}}
\nc\bbR{\mathbb{R}}
\nc\bbZ{\mathbb{Z}}

\nc{\todo}[1]{\textcolor{red}{todo: #1}}
\nc{\Anote}[1]{\textcolor{red}{Aram note: #1}}

\newcommand{\vect}[1]{\boldsymbol{#1}}

\newcommand{\bv}{{\vect{b}}}
\newcommand{\vgamma}{\vect{\gamma}}
\newcommand{\vbeta}{\vect{\beta}}

\def\begsub#1#2\endsub{\begin{subequations}\label{eq:#1}\begin{align}#2\end{align}\end{subequations}}
\nc\qand{\qquad\text{and}\qquad}
\nc\mnb[1]{\medskip\noindent{\bf #1}}

\nc{\pder}[2]{\frac{\partial {#1}}{\partial {#2}}}
\nc{\p}{\partial}
\title{The Quantum Approximate Optimization Algorithm Needs to See the Whole Graph: Worst Case Examples}
\author[1]{Edward Farhi}
\author[2]{David Gamarnik}
\author[ ]{Sam Gutmann}
\affil[1]{\small Google Inc., Venice CA 90291 and
Center for Theoretical Physics, MIT, Cambridge MA, 02139}
\affil[2]{\small Operations Research Center and Sloan School of Management MIT, Cambridge MA, 02140}

\begin{document}

\maketitle
\begin{abstract}
    The Quantum Approximate Optimization Algorithm can be applied to  search problems on graphs with a cost function that is a sum of terms corresponding to the edges.  When conjugating an edge term,  the QAOA unitary at depth p produces an operator that depends only on the subgraph consisting of edges that are at most p away from the edge in question.  On random d-regular graphs, with d fixed and with p a small constant time log n, these neighborhoods are almost all trees and so the performance of the QAOA is determined only by how it acts on an edge in the middle of tree. Both bipartite random d-regular graphs and general random d-regular graphs locally are trees so the QAOA's performance is the same on these two ensembles. Using this we can show that the QAOA with $(d-1)^{2p}  < n^A$ for any $A<1$,  can only achieve an approximation ratio of 1/2 for Max-Cut on bipartite random d-regular graphs for d large. For Maximum Independent Set, in the same setting, the best approximation ratio is a d-dependent constant that goes to 0 as d gets big.
    
\end{abstract}

\section{Introduction}
The Quantum Approximate Optimization Algorithm \cite{QAOA1} \cite{QAOA2} can be applied to finding approximate solutions to combinatorial optimization problems on graphs. Examples that we focus on in this paper are Max-Cut (MC) and Maximum Independent Set (MIS). The problem specific unitary operator used in the algorithm is local - it only interacts qubits connected by an edge in the graph. When the algorithm is run at depth $p$, the measurement outcomes of a given qubit depend only on the $p$-neighborhood of that qubit, that is, those qubits within a distance $p$ in the graph. If these neighborhoods are small, the QAOA does not ``see" the whole graph and in some cases there are known bounds on the algorithmic performance \cite{BKKT} \cite{FGamG}. In this paper we show that if $p$ is less than a certain multiple of $\log n$, the QAOA cannot come close to optimal when applied to bipartite random $d$-regular graphs. The trick here is to notice that bipartite random $d$-regular graphs locally look like trees just like general random $d$-regular graphs so at shallow depth the QAOA has the same performance on both.  However the two distributions have very different Max-Cut values and Maximum Independent Set sizes resulting in an obstacle to performance in the bipartite case.  

Our results show limitations on the performance of the QAOA when $(d-1)^{2p}$ is a vanishing fraction of the number of bits as $n$ gets large. Beyond this, when the QAOA ``sees" the whole graph we have no indication that performance is limited. Even with, say,  one million qubits at d=3 our results only show that the algorithm is limited when $p$ is less than $10$ which in practice can be viewed as shallow depth.

This paper is a companion to our recent paper \cite{FGamG} showing the limitation of the QAOA for finding big independent sets in random graphs with $\frac{n d}{2}$ edges with $d$ fixed.  That was a  ``typical" case result with a well known distribution.  Here we change the distribution and get a stronger result in the sense that the QAOA outputs strings that are further from optimal. We get this by looking at bipartite random $d$-regular graphs. With the previous distribution every vertex has average degree $d$ whereas with the new distribution every vertex has the same degree $d$.  The substantive difference is the restriction to bipartite graphs which we do not view as ``typical" and for that reason we call this a ``worst" case result.  Also in the previous work we used a property of large independent sets in random graphs, the Overlap Gap Property.  Here the only property of random $d$-regular graphs we use (both bipartite and general) is that locally they look like trees. Our techniques here are equally applicable to Max-Cut for which the Overlap Gap Property  is conjectured not to hold and to Maximum Independent Set where it is known to hold.

For problems defined on graphs the cost function can be written as
\begin{align}
    C= \sum\limits_{<ij>}\, C_{ij}
    \label{eq:costfunction}
\end{align}
where $C_{ij}$ is the cost function associated with the edge connecting vertices $i$ and $j$. If the bit values on the vertices are such that the edge constraint is satisfied then $C_{ij} = 1$, if it is not then $C_{ij} = 0$. The goal is to make $C$ big. 

For Max-Cut the associated cost function is
\begin{align}
    C^{\mathrm{MC}}_{\,ij} = b_i +b_j - 2b_i b_j~,
    \label{maxcut}
\end{align}
where each $b_i$ is 0 or 1. This cost is satisfied when the bits disagree. Note that every bit string can be associated with a cut value.

For Maximum Independent Set the choice of cost function is trickier. Here a generic bit string does not correspond to an independent set. So the cost function we use has two parts. One is the Hamming weight which we want to make big. The other penalizes strings that are not independent sets. In this paper we focus on $d$-regular graphs and we take
\begin{align}
   C^{\mathrm{MIS}}_{\,ij} = \frac{1}{2 d} \left(b_i + b_j\right) - b_i b_j ~.
   \label{hamming}
\end{align}
Note that since we are summing on edges in $d$-regular graphs the first term gives the Hamming weight in \eqref{eq:costfunction}. If the algorithm outputs a string ${\bf b} = b_1,\ldots,b_n$ with cost $C({\bf b})= C_{out} > 0$ then by pruning we can produce a string corresponding to an independent set of size at least $C_{out}$. To see this consider two bits that are $1$ sitting at vertices with an edge between them. Set one of these to $0$. This decreases the Hamming weight by $1$ but decreases the penalty by at least $1$ so the cost function does not go down. Continue in this way until all the penalty terms are satisfied and the resulting independent set has size at least $C_{out}$.
\section{Quantum Approximate Optimization Algorithm}

Let us review the ingredients of the QAOA.
The  graph-dependent cost function $C$ gives rise to  an operator that is diagonal in the computational basis, defined as
\begin{equation}
C\ket{\bv} = C(\bv)\ket{\bv}.
\end{equation}
The problem dependent unitary operator depends on $C$ and a single parameter $\gamma$,
\begin{align} \label{eq:UC}
U(C,\gamma) = e^{-i\gamma C}.
\end{align}
Note that $U(C,\gamma)$ conjugating a single qubit operator produces an operator that only involves that qubit and those connected to it on the graph.

The operator that induces transitions between strings uses
\begin{align}
B = \sum_{j=1}^n X_j,
\end{align}
where $X_j$ is the Pauli $X$ operator acting on qubit $j$, and the associated unitary operator  depends on a parameter $\beta$,
\begin{align} \label{eq:UB}
U(B,\beta) = e^{-i\beta B} = \prod_{j=1}^n e^{-i\beta X_j}.
\end{align} Note that $U(B,\beta)$ conjugating a single qubit rotates that qubit and has no effect on other qubits.

We initialize the system of qubits in a symmetric product state such as 
\begin{align} \label{eq:psi0}
\ket{s} = \ket{0}^{\otimes n} 
\end{align}
or
\begin{equation}
\ket{s} = \ket{+}^{\otimes n} =  \frac{1}{\sqrt{2^n}} \sum_{\bv}\ket{\bv}.
\end{equation}
Using a product state for the initial state is the usual choice for the QAOA and is required for the arguments below.

The quantum circuit alternately applies $p$ layers of $U(C,\gamma)$ and $U(B,\beta)$.
With $\vect\gamma = \gamma_1,\gamma_2,\ldots,\gamma_p$ and $\vect\beta = \beta_1,\beta_2,\ldots,\beta_p$ 
we have the unitary operator
\begin{align} \label{eq:unitary}
U=U(B,\beta_p) U(C,\gamma_p) \cdots U(B,\beta_1) U(C, \gamma_1)
\end{align}
which acting on the initial state gives 
\begin{align} \label{eq:wavefunction}
\ket{\vect\gamma, \vect\beta} = U\ket{s}.
\end{align}
The associated QAOA objective function is
\begin{align}
\bra{\vgamma,\vbeta}C\ket{\vgamma,\vbeta} \label{eq:QAOA_obj}
\end{align}
and the goal is to find $\vect\gamma$ and $\vect\beta$ to make this big. We will not concern ourselves with how optimal parameters are found and for our arguments we just assume we have them.

Using \eqref{eq:costfunction} we have
\begin{align} 
\bra{\vgamma,\vbeta}C\ket{\vgamma,\vbeta}=\sum\limits_{<ij>}\, \bra{s} U^\dag C_{ij} U \ket{s}
\label{eq:CisC}
\end{align}
and for our two examples the cost function has the same form on each edge. Because of the locality of $U(C,\gamma)$ we see that the right hand side depends only on the $p$-neighborhoods of each edge. By an edge-$p$-neighborhood we mean the subgraph that contains all of the edges within a distance $p$ of the edge being examined.
\section{Main Result}

We first outline how we get our worst case results. For a random $d$-regular graph, almost all of the edge-$p$-neighborhoods are trees as long as $(d-1)^{2 p} < n^A$ for some $A<1$. We will prove this shortly.  With $p$ small enough, almost every edge in \eqref{eq:CisC}  has a tree neighborhood and so the performance of the QAOA is governed by how it evaluates an edge clause in the middle of a tree.  The QAOA cannot do better than finding the optimal cost and on a random $d$-regular graph there are known bounds on the size of the biggest cut and the biggest independent set. These bounds then bound the quantum expectation of an edge clause in the middle of a tree. However on a bipartite graph Max-Cut is completely satisfiable and there is an independent set containing half of the vertices.  But in a bipartite random $d$-regular graph
almost all edges have $p$-neighborhoods that are trees. So the performance of the QAOA is the same on bipartite random $d$-regular  graphs as it is on  general random $d$-regular graphs.  This means that the QAOA, at shallow depth, will not come near the optimal for MC or MIS on bipartite random $d$-regular graphs. We now flesh out these arguments.

To begin we prove what we need about neighborhoods that are trees in random graphs.   Suppose that
\begin{align}
\label{eq:condition}
(d-1)^{2 p}<n^A ~~~ \mathrm{for~some}  ~~~ A<1 .
\end{align}
If the neighborhood of an edge is not a tree it contains a cycle. Any cycle in the $p$-neighborhood of an edge can not have length greater than $2 p + 1$. So we first quote some facts \cite{Trees} about the number of not very big cycles in random graphs. Consider a cycle of length $k=O(\log n)$ in a general or in a bipartite random $d$-regular graph. Then 
\begin{align}
    \mathrm{Ex}[ ~\textit{number of cycles of length} = k ~] = O\big((d-1)^k\big)
\end{align}
and  therefore
\begin{align}
\mathrm{Ex}[ ~\textit{number of cycles of length} \le k ~] = O\big((d-1)^k\big).
\end{align}
From this it follows that for any $\epsilon > 0$
\begin{align}
\label{eq:Prob}
    \mathrm{Prob}\big[~(\textit{number of cycles of length} \le k) \ge (d-1)^k n^\epsilon ~\big] \rightarrow 0 ~~~\text{as}~~~ n \rightarrow \infty.
\end{align}
We will take $\epsilon < 1 - A$.
Here Ex and Prob are with respect to the graph distributions.  

We say that a cycle ``ruins" an edge if the cycle sits in the $p$-neighborhood of the edge. We call the edge whose $p$-neighborhood we are looking at the middle edge. If the cycle contains the middle edge, the cycle length is bounded by $2 p + 1$.  If the $p$-neighborhood of the middle edge contains a cycle whose closest edge is $\Delta$ from the middle then the cycle length is $\le (2 p + 2 - 2 \Delta)$. We now consider a cycle and ask how many edges, whose closest distance to the cycle is $\Delta$, it can ruin. Focus on any  edge  in the cycle. At a distance $\Delta$ from this edge there are at most $2 (d-1)^\Delta$ other edges.  Each of these can be viewed as a middle edge that might be ruined by the cycle. So there are at most $2 (d-1)^\Delta (2 p + 2 -2 \Delta) $ middle edges that can be ruined by the cycle at distance $\Delta$. Now from \eqref{eq:Prob} there are at most $O((d-1)^{2p+2-2\Delta}~ n^\epsilon)$ such cycles so at most $O(p (d-1)^{2 p + 2 - \Delta}~n^\epsilon)$ middle edges can be ruined for this $\Delta$.  Summing over $\Delta$ we have at most $O(p^2 (d-1)^{2p}~n^\epsilon)$ edges in the graph that can be ruined.  Let $A'$ be greater than $A+\epsilon$ but less than $1$ and we have that
\begin{align}
    \mathrm{Ex}[~\textit{number of edges that are not trees}~] = O(n^{A'}).
\end{align}

We now turn to the performance of the quantum algorithm. Focus on edges that have $p$-neighborhoods that are trees and run the algorithm to depth $p$.
For Max-Cut let
\begin{align}
    \textbf{C}^{\mathrm{MC}}_{\mathrm{tree}}=\bra{s} U^\dag C^{\mathrm{MC}}_{ij} U \ket{s}
    \label{eq:lamMC}
\end{align}
and for Maximum Independent Set let
\begin{align}
    \textbf{C}^{\mathrm{MIS}}_{\mathrm{tree}}=\bra{s} U^\dag C^{\mathrm{MIS}}_{ij} U \ket{s}
    \label{eq:lamMIS}
\end{align}
where the edge $ij$ has a $p$-neighborhood that is a tree. These are the contributions to the quantum expectation of the cost function from single edges that have tree neighborhoods. 
For random $d$-regular graphs almost every edge has a tree neighborhood so we have
\begin{align}
    \mathrm{Ex}\big[\bra{s} U^\dag C^{\mathrm{MC}} U \ket{s}\big]= \frac{n d}{2} \textbf{C}^{\mathrm{MC}}_{\mathrm{tree}} + O(n^{A'})
\end{align}
where again $A'<1$ and the expectation is with respect to the graph distribution. The  $O(n^{A'})$ accounts for the fact that not all edges have tree neighborhoods and that the QAOA may not give \eqref{eq:lamMC} on these edges. Similarly for MIS on random $d$-regular graphs
\begin{align}
    \mathrm{Ex}\big[\bra{s} U^\dag C^{\mathrm{MIS}} U \ket{s}\big]= \frac{n d}{2} \textbf{C}^{\mathrm{MIS}}_{\mathrm{tree}} + O(n^{A'}).
\end{align}

We now state some established results about MC and MIS on random $d$-regular graphs.  For Max-Cut there exists a $\rho_d$ such that the optimal cut is
\begin{align}
    \rho_d~n + o(n)
\end{align}
with high probability for large $n$. 
For Maximum Independent Set there exists a $\sigma_d$ such that the biggest independent set on typical $d$-regular graphs is
\begin{align}
    \sigma_d~n + o(n)
\end{align}
with high probability for large $n$. 
Since no algorithm, including the QAOA, can ever do better than the optimum we have
\begin{align}
\label{eq:MCle}
\textbf{C}^{\mathrm{MC}}_{\mathrm{tree}}\le \frac{2 \rho_d}{d}
\end{align}
and
\begin{align}
\label{eq:MISle}
\textbf{C}^{\mathrm{MIS}}_{\mathrm{tree}}\le \frac{2 \sigma_d}{d}.
\end{align}

Let us now switch to bipartite random $d$-regular graphs.  Since even in this case  essentially all of the $\frac{n d}{2}$ edges have tree neighborhoods we can use \eqref{eq:lamMC} and \eqref{eq:lamMIS} and multiply by  $\frac{n d}{2}$ to get the total quantum cost.  Then using \eqref{eq:MCle} and \eqref{eq:MISle} we get for bipartite random $d$-regular graphs
\begin{align}
    \mathrm{Ex}\big[\bra{s} U^\dag C^{\mathrm{MC}} U \ket{s}\big]\le
\rho_d~n + O(n^{A'})
\end{align}
and
\begin{align}
    \mathrm{Ex}\big[\bra{s} U^\dag C^{\mathrm{MIS}} U \ket{s}\big]\le
\sigma_d~n  + O(n^{A'}).
\end{align}
However any bipartite $d$-regular graph has a cut of size $\frac{n d}{2}$. This means that the QAOA can not produce a cut on a bipartite random $d$-regular graph with an approximation ratio (output value over best possible) better than
\begin{align}
\frac{2 \rho_d}{d} + O(n^{A'-1}).
\end{align}
It is known \cite{3MC}  that $\rho_3 < 1.4026$. It is also known \cite{MCbigd}\cite{Dembo} that for $d$ large $\rho_d\le\frac{d}{4}+O(\sqrt{d})$ so the approximation ratio is no better than $\frac{1}{2}$ as $d$ grows big.

Also any bipartite $d$-regular graph has an independent set of size $\frac{n}{2}$ so the QAOA can not produce an independent set with an approximation ratio better than
\begin{align}
2 \sigma_d + O(n^{A'-1}).
\end{align}
It is known \cite{cubic} that $\sigma_3<.454$. We also know  \cite{bigd} that for $d$ large $\sigma_d\le\frac{2 \ln d}{d}$ so the approximation ratio is going to $0$ as $d$ grows big.
\section{Discussion}
Given a graph problem on a bounded degree graph, the QAOA with the usual form for the operators $B$ and $C$, is a local quantum algorithm.  The unitary applied at each level of the QAOA, when conjugating a single qubit operator produces an operator only involving that  qubit and those connected to it on the graph. This means that at sufficiently shallow depth the QAOA does not ``see" the whole graph.  The QAOA acting on two graphs which locally look alike will perform the same on the two graphs. But two graphs can locally look alike and still be different when viewed as wholes. For combinatorial search problems on graphs, large scale structure can affect the optimum.  We exploit this to construct examples where the QAOA's performance is provably below optimal.  We look at bipartite random $d$-regular graphs.  For Max-Cut when $p$ is a small enough constant times $\log n$  we show that the approximation ratio is no better than $\frac{1}{2}$ for large $d$.  (Under a symmetry assumption, the worst case example of \cite{BKKT} for Max-Cut bounds the approximation ratio by a $d$-dependent constant that is bigger than  $\frac{5}{6}$ for all $d$.)  For Maximum Independent set the approximation ratio goes to $0$ at large $d$. However our results only apply at shallow depth, that is, with $p$ growing no faster than a  constant times $\log n$.   At higher depth the QAOA sees the whole graph, are techniques do not apply, and we need to see how close to optimal the QAOA can get.

\bibliographystyle{unsrt}
\bibliography{bibliography}

\end{document}